\documentclass[a4paper]{spie}  

\usepackage[]{graphicx}

\title{Competitive Advantage for Multiple-Memory Strategies \\
in an Artificial Market} 

\author{Kurt E. Mitman, Sehyo Charley Choe,  and Neil F. Johnson
\skiplinehalf
Clarendon Laboratory, Physics Department, Oxford University, Oxford OX1 3PU, U.K. 
}

\authorinfo{Further author information: (Send correspondence to K.E.M.)
\\
K.E.M.: E-mail: k.mitman1@physics.ox.ac.uk}

\begin{document} 
  \maketitle 

\begin{abstract}
We consider a simple binary market model containing $N$ competitive agents. The novel feature of our model is 
that it incorporates the tendency shown by traders to look for patterns in past price movements over multiple 
time scales, i.e. {\em multiple memory-lengths}. In the regime where these memory-lengths are all small, the 
average winnings per agent exceed those obtained for either (1) a pure population where all agents have equal 
memory-length, or (2) a mixed population comprising sub-populations of equal-memory agents with each 
sub-population having a different memory-length. Agents who consistently play strategies of a given 
memory-length, are found to win more on average -- switching between strategies with different memory lengths 
incurs an effective penalty, while switching between strategies of equal memory does not. Agents employing 
short-memory strategies can outperform agents using long-memory strategies, even in the regime where an 
equal-memory system would have favored the use of long-memory strategies. Using the many-body 
`Crowd-Anticrowd' theory, we obtain analytic expressions which are in good agreement with the observed 
numerical results.  In the context of financial markets, our results suggest that multiple-memory agents have 
a better chance of identifying price patterns of unknown length and hence will typically have higher winnings. 
\end{abstract}



\section{INTRODUCTION}
\label{sect:intro}  
Complex systems are thought to be ubiquitous in the physical, biological and economic world. Research across 
these disciplines has focused increasingly on complex adaptive systems and their dynamical behavior.  A noted 
feature of complex adaptive systems is that they can give rise to large macroscopic changes or `extreme 
events' that appear spontaneously, have long-lasting consequences and yet seem to be very difficult to 
predict.  Recent research \cite{newsci,lamper,sornette} indicates, however, that predictability may be 
possible in certain instances.  The models employed in such studies have typically involved individual, 
self-interested agents competing for a limited resource \cite{newsci,lamper,sornette}.  Research on these 
limited-resource agent games has so far been limited to `pure' populations 
\cite{agents,NFJ1,TMG,netMG,netMG2,spy} where all agents have the same memory-length \emph{m}, and `alloy' 
populations where the population of agents consists of sub-populations of equal $m$ but where $m$ varies from 
sub-population to sub-population \cite{alloy}.

In this paper we examine the effects of multiple-memory strategies in a multi-agent population. In particular, 
we consider the situation where the strategy set of an individual  agent contains strategies with different 
memory-lengths \emph{m}.  When the memory-lengths are sufficiently small, these multiple-memory agents 
outperform both a pure population of equal-memory agents, and an alloy population comprising sub-populations 
of equal-memory agents.  Agents who consistently play strategies of a given memory length, are found to win 
more on average -- switching between strategies with different memory lengths incurs an effective penalty, 
while switching between strategies of equal memory does not. We find that agents choosing to use short-memory 
strategies can outperform agents using long-memory strategies -- remarkably, this is  true even in the regime 
where an equal-memory system would have favored the use of long-memory strategies. Using the many-body 
`Crowd-Anticrowd' theory \cite{NFJ1} we obtain analytic expressions which are in good agreement with the 
observed numerical results.  In the context of financial markets, our results suggest that multiple-memory 
agents have a better chance of identifying price patterns of unknown length and hence will typically have 
higher winnings. In other words, agents who are capable of looking for patterns in past outcomes over several 
timescales, will do better on average.

Our artificial market takes the form of a binary agent resource (B-A-R) game \cite{NFJ1,arthur} of which a 
special limiting case is the so-called Minority Game\cite{agents,NFJ1}.  The market consists of an odd number 
of agents (e.g. traders) $N$, of which no more than $L<N$ agents can be rewarded at each time step.  At each 
time step $t$ each agent $i$ makes a decision ${a_i}(t)$ to buy (${a_i}(t)=1)$ or sell (${a_i}(t)=0$).  If 
$\sum_{i=1}^N {a_i}(t) > L$ then at time step $t$ all agents who chose `0' are rewarded, otherwise all agents 
who chose `1' are rewarded.  As a specific example in this paper, we take \emph{N} to be odd, and set 
$L=(N-1)/2$ which results in more losers than winners at each time step as in the Minority Game. Elsewhere we 
give the corresponding results for general $L$.  This artificial market  provides a simple paradigm for the 
dynamics underlying financial markets: more sellers than buyers implies lower prices, thus it can be better 
for a trader to be in the smaller group of buyers.  The `output' of the market is a single binary digit, 0 or 
1 (0 if $\sum_{i=1}^N {a_i}(t) > L$, otherwise 1).  This is the \emph{only} information that is available to 
the agents.  Agents with strategies of memory size \emph{m} therefore have access to the last \emph{m} binary 
output digits $\mu_m$ when making their decisions.  Agents who possess multiple strategies of different memory 
lengths will therefore use different information when deciding which action to take at the next time step.  
Using information from different history lengths can be interpreted as somewhat analogous to the techniques of 
`chartists' for making forecasts. Indeed, it is well known that in practice a financial trader's computer 
screen  displays past price movements over several different timescales (e.g. hours, days, weeks).  

We consider our \emph{N}-agent population to possess strategies drawn from a strategy pool which corresponds 
to memory lengths $m_1$ and $m_2$.  At the beginning of the game the agents are randomly assigned $s$ 
strategies, of which $s_1$ strategies are of memory length $m_1$ and $s_2$ strategies are of memory length 
$m_2$ with repetitions allowed. The total number of strategies for each agent is the same, i.e. $s = s_1 + 
s_2$. A strategy is a mapping from the length $m$ recent-history bit-string onto a binary decision 
$\mu_m\rightarrow \{0,1\}$.  Consider $m=3$. There are $2^{2^m} = 256$ possible strategies, each of which can 
be represented by a string of 8 bits (0 or 1) corresponding to the decisions based on the $2^m = 8$ possible 
histories $\mu_m$ (e.g. 000, 001, etc).  For example, the $m=3$ strategy $01101000$ represents the mapping 
$\{000\rightarrow 0, 001\rightarrow 1, 010\rightarrow 1, 011\rightarrow 0, 100\rightarrow 1, 101\rightarrow 0, 
110\rightarrow 0, 111\rightarrow 0\}$.

  On every time step, the agent is awarded one point if the strategy he chose to use does actually predict the 
correct global output.  In addition, he compares the prediction of all of his strategies to the correct global 
output.  Each strategy that predicted the correct global output receives one `virtual' point, and each 
strategy that predicted the wrong global output loses one `virtual' point.  Thus at each time step, each agent 
has a running tally of how successful each of his \emph{s} strategies has been.  On each time step an agent 
picks the most successful strategy (with the highest `virtual' point score) as his decision.  If two or more 
strategies have the same `virtual' point score the agent randomly chooses between the tied strategies with 
equal probability.  The success of any particular strategy generally fluctuates.  As agents begin to use 
similar strategies, those strategies become less profitable, causing the agents to switch to a different 
strategy.  Therefore there is no best strategy for all times.

The full strategy space (FSS) forms a $2^m$-dimensional hypercube for memory length $m$ with a unique strategy 
at each of the $2^{2^m}$ verticies \cite{agents,NFJ1}.  In general, the game's dynamics can be reproduced by 
considering a reduced strategy space (RSS) containing only $2\times 2^m$ strategies, where each strategy is 
either anti-correlated or uncorrelated to the rest of the strategies in the RSS \cite{rss}. This reduction of 
the FSS to RSS has the effect of retaining the strong correlations in the system, which in turn tend to drive 
the dynamics, while removing the weak ones which just tend to create minor fluctuations. If the total number 
of strategies in play is greater than the size of the reduced strategy space (i.e. $N\times s \gg 2\times 
2^m$) many agents may hold the highest-scoring strategy at any given time step.  This will cause a large 
number of agents to play the same strategy, which will lead to a large disparity between the number of agents 
who chose winning and losing market actions (i.e. $\sum_{i=1}^N {a_i}(t) \gg L$ or $\sum_{i=1}^N {a_i}(t) \ll 
L$) and a relatively low number of total points awarded \cite{agents,NFJ1} since the total number of points 
awarded at each time step is given by $(N-{\rm abs}[\sum_{i=1}^N {a_i}(t)-L-0.5]-0.5)/2$.   Such crowd effects 
are a strategy-space phenomenon and have been shown to quantitatively explain the fluctuations in the number 
of winning agents per turn for the pure population as a function of $m$ and $s$ \cite{NFJ1}.  Furthermore, the 
most number of points that can be awarded on a given time step is $(N-1)/2$ and hence the average winnings 
(total points awarded) per agent per turn, $W$, is always less than or equal to $(N-1)/2N$, hence $W<0.5$.  
When $N\times s \gg 2\times 2^m$, $W$ is substantially less than 0.5 due to the crowd effects mentioned above.  
Note that an external (i.e. non-participating) gambler using a `coin toss' to predict the winning room, would 
have a $50\%$ success rate since he would not suffer from this intrinsic crowding in strategy space.

The dynamics of the artificial market also depend on the trajectory which the game follows in the history 
space $\{\mu\}$.  The history space forms an \textit{m}-dimensional hypercube whose $2^m$ vertices correspond 
to all possible history bit-strings of length $m$.  In the crowded regime (i.e. $N\times s \gg 2\times 2^m$) 
there is information left in the history time series that is inaccessible by a pure population of agents with 
memory $m$, since this information is contained within bit-strings of length greater than $m$.  When the total 
number of strategies held by the agents is only a small subset of the RSS, agents are not able to fully 
extract the information contained in bit-strings of any length (including $\leq m$) \cite{NFJ1,alloy}.  
Cavagna claimed \cite{cavagna} that it is irrelevant whether one uses the real history or random history as an 
input to the agents in the Minority Game \cite{cavagna}.
However Johnson \emph{et al.} subsequently showed \cite{alloy} that in an alloy population of agents, for 
example, the trajectories through history space do indeed become important. In particular, higher $m$ agents 
have the opportunity to exploit specific correlations in the real-history time series left by the lower $m$ 
agents.

  \begin{figure}
   \begin{center}
   \begin{tabular}{c}
   \includegraphics[height=7cm]{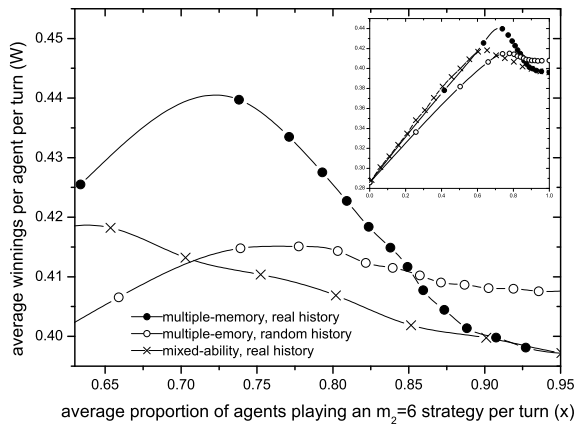}
   \end{tabular}
   \end{center}
   \caption[fig1] 
   { \label{fig:fig1} 
Average winnings per agent per turn, $W$, for multiple-memory populations and a mixed-ability population with 
$N=101$, $m_1=3$ and $m_2=6$, as obtained numerically.  Each agent has $s=16$ strategies.  Each data-point 
corresponds to an average over 25 simulations. The connecting curves are a guide to the eye.  The inset shows 
$W$ over the full range of $x$, where $x$ is the average proportion of agents playing an $m_2=6$ strategy per 
turn. See the text for an explanation of `real history' and `random history'.}
   \end{figure} 

\section{RESULTS}

Figure 1 shows the average winnings per agent per turn, $W$, for two multiple-memory populations (circles) and 
a mixed-ability population (crosses).  All three populations correspond to $N=101$ agents and $s=16$ 
strategies per agent.  
In the mixed-memory population, $N_{m_1}$ agents hold strategies of memory-length $m_1 = 3$, while $N-N_{m_1}$ 
agents hold strategies of memory-length $m_2=6$.  In the multiple-memory populations, each agent holds $s_1$ 
strategies of memory-length $m_1=3$ and $s_2$ strategies of memory-length $m_2 = 6$.  The curves are plotted 
against $x$, the average proportion of agents playing an $m_2$ strategy on each turn.  For the mixed-ability 
population, the average proportion of agents playing an $m_2$ strategy is simply $(N-N_{m_1}) / N$.  For the 
multiple-memory populations, we determined numerically the average number of agents playing an $m_2$ strategy 
at each turn, and divided that number by $N$.  Each data point represents an average over 25 runs.   The 
results for the multiple-memory population are averaged over both $W$ and $x$, since in the multiple-memory 
population $s_1$ and $s_2$ are exogenously determined whereas $x$ is endogenous (see Figure 2).  For clarity 
we have not shown the error bars for each data point -- however the range of values in both $W$ and $x$ is 
sufficiently small that our results, discussions and conclusions are not affected by numerical 
artifacts\footnote{NB: the spread in $x$ for mixed-ability populations is zero.}.  Agents are supplied with 
the $m_1$ (and/or $m_2$) most recent winning actions of the artificial market in the `real-history' results 
(solid circles for the multiple-memory population).  For the `random-history' results (open cirles for the 
multiple-memory population) agents are given a random $m_1$ (and/or $m_2$) length bit-string every time step 
instead of the actual bit-string of winning market actions. (For random-history results for a mixed-ability 
population, see Johnson \emph{et. al.}\cite{alloy}).  The `real-history' multiple-memory population exhibits a 
maximum in $W$ at a finite value of $x\sim .72$.  We find that as $s$ increases the value of $x$ that 
maximizes $W$ asymptotically approaches $\sim .76$ (we have investigated cases up to $s=100$).   The total 
number of points awarded per turn can therefore exceed either a pure $m_1=3$ or pure $m_2=6$ population of 
agents.  The `random-history' multiple-memory population also exhibits a maximum at a finite value of $x$.  
However, the magnitude of the difference between $W_{\rm{max}}$ and $W$ when $x=1$ (i.e. a pure $m_2$=6 
population) is only $\sim 0.007$, which is within the standard deviation of the values of $W$. Hence the 
`random-history' multiple-memory population does not significantly outperform a pure $m_2=6$ population.  
Therefore, history has a significant effect on the ability of a multiple-memory population to outperform a 
pure population.  In other words, {\em a multiple-memory population can collectively profit from the real 
patterns which arise in past outcomes -- most importantly, this added benefit of multiple-memory does {\bf 
not} arise simply from a reduction in crowding in the strategy space}. For all values of $x$, the average 
winnings per agent per turn of the `real-history' multiple-memory population, is greater than or equal to the 
average winnings per agent per turn for the mixed-ability population.  This result is not duplicated if the 
multiple-memory population is presented with a random history string.  Multiple-memory agents that are 
presented with the real history of winning actions from the artificial market can therefore outperform both 
pure populations of agents holding strategies of memory length $m_1$ or $m_2$ \emph{and} simultaneously 
outperform a mixed-ability population consisting of sub-populations of agents with $m_1$ or $m_2$ strategies.

   \begin{figure}
   \begin{center}
   \begin{tabular}{c}
   \includegraphics[height=7cm]{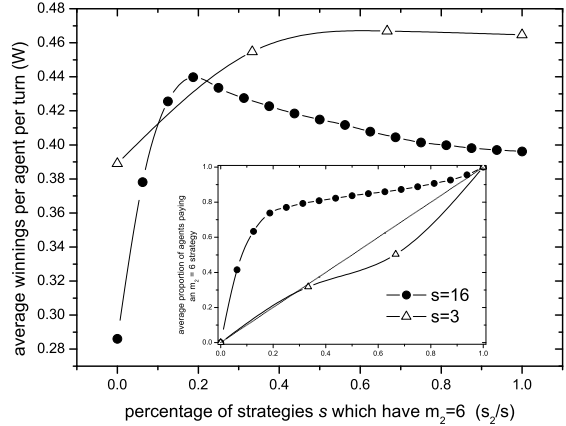}
   \end{tabular}
   \end{center}
   \caption[fig2] 
   { \label{fig:fig2} 
Average winnings per agent per turn, $W$, for multiple-memory populations with $s=16$ and $s=3$.  Other 
parameters are as in Fig. 1.  The inset shows the average proportion $x$ of agents playing an $m_2$ strategy  
for the same two populations.  The dotted line in the inset represents the line $x=s_2/s$.}
   \end{figure} 

Figure 2 shows the average winnings per agent per turn $W$ for two multiple-memory populations.  Both 
populations are given the real history bit-string. One population has $s=16$ (solid circles) as above, while  
the other population has $s=3$ (triangles).  All other relevant parameters are the same as in Fig. 1, however 
we now plot $W$ against the exogenously controlled proportion of $m_2$ strategies (i.e. $s_2/s$).  Both $s=16$ 
and $s=3$ multiple-memory populations exhibit a maximum in $W$ at finite $s_2/s$, again showing that the 
multiple-memory populations can outperform pure populations.  As $s$ increases, the value of $s_2/s$ which 
maximizes $W$ tends towards 0. However, for $7\leq s\leq 100$ we find that $W$ is always maximized when 
$s_2=3$.  We find a similar result for multiple-memory populations with $m_1=1$ and $m_2=2$ -- in this case, 
$W$ is always maximized when $s_2 = 1$.  In the inset to Fig. 2, we plot the average proportion of agents who 
play an $m_2=6$ strategy per turn (i.e. $x$) versus $s_2/s$.  The circles and triangles represent the same 
multiple-memory populations as described above, and the straight dashed line corresponds to $x=s_2/s$.  
Neglecting the endpoints, $x$ for the $s=16$ multiple-memory population always lies above the dashed line.  
Agents are therefore playing their $m_2$ strategies at a higher rate than if they were choosing a strategy at 
random each turn.  This result is not too surprising, since we expect the $m_2$ strategies to outperform the 
$m_1$ strategies -- after all, a pure $m_2$ population will have a higher $W$ than the corresponding pure 
$m_1$ population.  However when $s=3$ (or $s=2$), and neglecting the endpoints, $x$ lies below the dashed 
line.  Therefore in a multiple-memory population with low $s$, agents play $m_1$ strategies more frequently 
than the proportion of $m_1$ strategies that they hold. This result is remarkable since a pure $m_2=6$, $s=3$ 
population is known to outperform a pure $m_1=3$, $s=3$ populations.  The agents who play $m_1$ strategies 
more than half the time (i.e. agents with $x<0.5$) have higher average winnings per turn than agents who play 
$m_2$ strategies more than half the time.  This is illustrated in Fig. 3.

     \begin{figure}
   \begin{center}
   \begin{tabular}{c}
   \includegraphics[height=7cm]{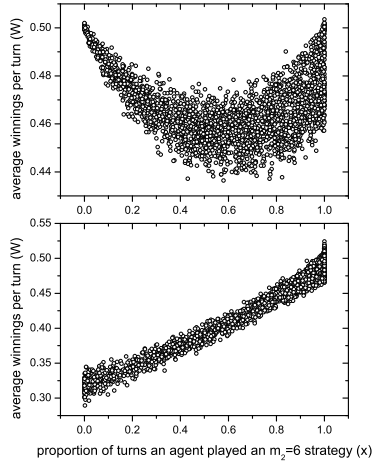}
   \end{tabular}
   \end{center}
   \caption[fig3] 
   { \label{fig:fig3} 
Average winnings per turn, $W$, for multiple memory populations with $s=3$ (top) and $s=16$ (bottom).  Each 
data-point represents a single agent. Other parameters are as in Fig. 2.}
   \end{figure} 

In the top half of Fig. 3, we plot the average winnings per turn for 2525 agents (25 runs, $N=101$ each run) 
versus the proportion of turns in which an agent played an $m_2$ strategy.  Each agent has $s=3$ strategies, 
with $s_1 = 1$, $m_1 = 3$, $s_2=2$ and $m_2=6$.  Agents who consistently play an $m_1$ or $m_2$ strategy have, 
on average, higher average winnings than agents who play a combination of $m_1$ and $m_2$ strategies.  In a 
pure population, agents who play a combination of their strategies also tend to incur an effective penalty.  
However, we find that the penalty for switching between strategies of different memory lengths is greater and 
more certain, i.e. $W$ is on average lower and the spread of $W$ values is also smaller.  In the bottom half 
of Fig. 3, we plot the average winnings per turn for 2525 agents  versus the proportion of turns in which an 
agent played an $m_2$ strategy.  All parameters are the same as for the top figure, except now we have set 
$s=16$, with $s_1=13$ and $s_2=3$.  With increasing $s$ it is clear that there is an effective penalty for 
playing an $m_1$ strategy.  Agents who consistently play $m_2$ strategies achieve winning percentages higher 
than 0.5, or that which could be achieved by an external player using a random coin toss to predict the 
winning market decision.  The average winnings per turn for agents who always play $m_2$ strategies is $\sim 
0.5$.

Figure 4 shows the standard deviation in the excess demand for a multiple-memory population (solid circles) as 
a function of the percentage of strategies which have $m_2=6$.  The parameters for the populations are the 
same as for the $s=16$ multiple-memory population discussed above.  The excess demand for our artificial 
market is the difference between the number of agents who choose to `buy' and `sell' at each time step.  The 
closer the excess demand is to zero, the higher the number of total points which are awarded each turn.  The 
standard deviation in the excess demand can serve as a proxy for the wastage in the system.  The more the 
standard deviation of excess demand fluctuates each turn, the smaller the total number of  points that can be 
awarded to agents.  The population exhibits a minimum in the standard deviation of excess demand at finite 
$s_2/s$, and at exactly the same value of $s_2/s$ which maximizes the average winnings per agent per turn (see 
Fig. 2).  The empty circles represent the standard deviation of the excess demand as predicted by the 
Crowd-Anticrowd theory, which is discussed in the following section.  In the inset we plot the standard 
deviation in excess demand for an $s=8$ multiple-memory population, with all other parameters being the same 
as in the main figure.

      \begin{figure}
   \begin{center}
   \begin{tabular}{c}
   \includegraphics[height=7cm]{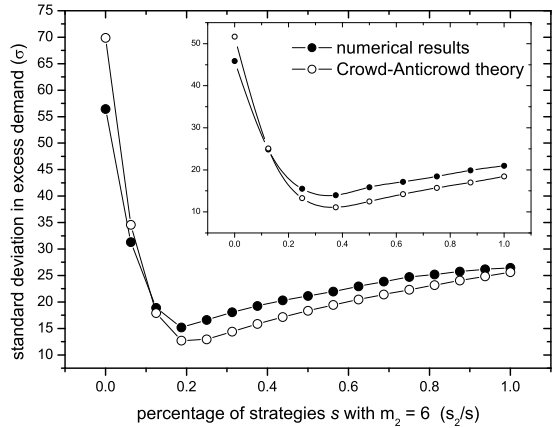}
   \end{tabular}
   \end{center}
   \caption[fig4] 
   { \label{fig:fig4} 
Standard deviation in the excess demand $\sigma$ for multiple memory populations with $s=16$ and $s=8$ (inset) 
obtained numerically (solid circles) and as predicted by the Crowd-Anticrowd theory (empty circles).  Other 
parameters are as in Fig. 1.  See text for an explanation of excess demand and Crowd-Anticrowd theory.}
   \end{figure}

\section{DISCUSSION}

Our numerical results demonstrate that populations of agents with multiple-memory strategies can outperform 
both pure populations of agents and mixed-ability populations.  This comparative advantage can be explained 
through the framework of the Crowd-Anticrowd theory.  As a first approximation, we treat the two groups of 
agents playing $m_1$ and $m_2$ strategies on a given turn as independent, as per the mixed-ability population.  
Thus we examine the Crowd-Anticrowd theory as applied to a mixed-ability population.  Considering the action 
of different sub-populations of agents as uncorrelated, the variance in the excess demand for a mixed-ability 
population goes as $\sigma^2 = \sigma_1^2 + \sigma_2^2$.  Here $\sigma_1^2$ ($\sigma_2^2$) is the variance due 
to the population of $m_1$ ($m_2$) agents, where $\sigma_1^2 = C^2(m_1,s_1)(1-x)^2 N^2$ and $\sigma_2^2 = 
C^2(m_2,s_2)(1-x)^2 N^2$. The pre-factor $C^2(m_i,s_i)$ is a constant of proportionality, and $x$ is the 
proportion of agents playing an $m_2$ strategy.  The standard deviation in excess demand can thus be 
calculated as:
\begin{equation}
\sigma = N[C^2(m_1,s_1)(1-x)^2+C^2(m_2,s_2)x^2]^{1/2}.
\label{sigma}
\end{equation}
For mixed-ability populations $x$ is exogenously determined. However in the multiple-memory population, $x$ is 
determined by the relative success of strategies with different memory-lengths.  Therefore we must develop an 
expression for how agents choose to segregate themselves between playing the $m_1$ and $m_2$ strategies.

In order to understand how agents will choose between $m_1$ and $m_2$ strategies, we must consider how the 
strategy spaces are related.  Every strategy in the $m_1$ space maps uniquely to a strategy in the $m_2$ 
space.  For example, take $m_1=3$, $m_2=6$ and the $m_1$ strategy $01101000$.  This strategy is equivalent to 
the $m_2$ strategy $0110100001101000011010000110100001101000011010000110100001101000$\footnote{representing 
the mapping
 $\{000000\rightarrow 0,
000001\rightarrow 1,
000010\rightarrow 1,
000011\rightarrow 0,
000100\rightarrow 1,
000101\rightarrow 0,
000110\rightarrow 0,
000111\rightarrow 0,
001000\rightarrow 0,
001001\rightarrow 1,
001010\rightarrow 1,
001011\rightarrow 0,
001100\rightarrow 1,
001101\rightarrow 0,
001110\rightarrow 0,
001111\rightarrow 0,
010000\rightarrow 0,
010001\rightarrow 1,
010010\rightarrow 1,
010011\rightarrow 0,
010100\rightarrow 1,
010101\rightarrow 0,
010110\rightarrow 0,
010111\rightarrow 0,
011000\rightarrow 0,
011001\rightarrow 1,
011010\rightarrow 1,
011011\rightarrow 0,
011100\rightarrow 1,
011101\rightarrow 0,
011110\rightarrow 0,
011111\rightarrow 0,
100000\rightarrow 0,
100001\rightarrow 1,
100010\rightarrow 1,
100011\rightarrow 0,
100100\rightarrow 1,
100101\rightarrow 0,
100110\rightarrow 0,
100111\rightarrow 0,
101000\rightarrow 0,
101001\rightarrow 1,
101010\rightarrow 1,
101011\rightarrow 0,
101100\rightarrow 1,
101101\rightarrow 0,
101110\rightarrow 0,
101111\rightarrow 0,
110000\rightarrow 0,
110001\rightarrow 1,
110010\rightarrow 1,
110011\rightarrow 0,
110100\rightarrow 1,
110101\rightarrow 0,
110110\rightarrow 0,
110111\rightarrow 0,
111000\rightarrow 0,
111001\rightarrow 1,
111010\rightarrow 1,
111011\rightarrow 0,
111100\rightarrow 1,
111101\rightarrow 0,
111110\rightarrow 0,
111111\rightarrow 0 \}$.}.  In the Crowd-Anticrowd theory, we assume that on each time step the ranking of 
strategies according to success rate and popularity are equivalent.  As there is a one-to-one mapping from 
$m_1$ strategy space to $m_2$ strategy space, we will assume that the relative rankings are also preserved in 
the mapping from $m_1$ space to $m_2$ space, i.e. if strategy $A_{m_1}$ is more popular than $B_{m_1}$, then 
we assume that $A_{m_2}$ is more popular than $B_{m_2}$).  Next we assume that an $m_2$ strategy will, with 
probability $p$, have a higher ranking than an agent's best $m_1$ strategy.  Therefore, the agent will play an 
$m_2$ strategy with probability
\begin{equation}
x = 1 - (1-p)^{(s-s_1)},
\label{x}
\end{equation}
where $s$ is the total number of strategies an agent possesses, and $s_1$ is the number of $m_1$ strategies 
that the agent possesses.  In order to determine the value for $p$, we must first calculate the expected value 
of the ranking $k$ for the agent's highest-ranked $m_1$ strategy as a function of $s_1$ and $m_1$:
\begin{equation}
E[k_{\rm{max}} | s_1] = 2P_1 ( 1 - {s_1\over s_1+1} )
\label{expected}
\end{equation}
where $P_1 = 2^{m_1}$.  If we analyze the strategies in terms of the RSS, the $m_1$ strategies ranked from 1 
to $P_1$ must map to $m_2$ strategies in the ranked set ${1..P_2}$.  This is a consequence of the fact that 
every strategy in the RSS is either anticorrelated or uncorrelated to every other strategy in the RSS. If both 
strategy spaces are in the crowded regime (i.e. $Ns_1 \gg 2P_1$ and $Ns_2 \gg 2P_2$) and the $m_2$ strategy 
space is significantly larger than the $m_1$ strategy space (i.e. $P_2/P_1 \gg 1$) then the mapping of the 
ranking of $m_1$ strategies will fall into the middle range of strategy-rankings of $m_2$ space. (This 
assumption should hold if $Ns_1/2P_1 \gg Ns_2/2P_2$).  For example, since ordering is preserved, the strategy 
with $k=1$ in the $m_1$ RSS maps to $k = P_2 - P_1 + 1$ in $m_2$ RSS.  Thus the probability that an $m_2$ 
strategy is better than the current most popular $m_1$ strategy, is $1 - [(P_2-P_1+1)/2P_2] = 
(P_2+P_1-1)/2P_2$.  More generally, the best $m_1$ strategy that an agent possesses is the $k_{\rm{max}}$th 
most popular one, given by Equation 3.  The general expression for $p$ then becomes 
\begin{equation}
p = {P_2 + P_1(1 - s_1/(s_1+1))-1\over 2P_2}.
\label{p}
\end{equation}

Our theoretical predictions for the standard deviation of the excess demand are plotted in Fig. 4.  The 
agreement with the numerical results is very good.  We also note that this agreement actually {\em improves} 
with increasing $s$, a feature that would be very hard to reproduce in comparable spin-glass based theories .  

One of the limitations of our theory as outlined so far, is the assumption that the actions of the agents 
playing an $m_1$ strategy is uncorrelated to the actions of agents playing an $m_2$ strategy.  As discussed 
above, the $m_2$ RSS covers the $m_1$ RSS -- therefore we need to modify Eq. 1 by adding a covariance term.  
The specific details of the covariance term will be presented elsewhere. For now, we just comment on the fact 
since there is likely to be additional crowding that is unaccounted for, this covariance should be positive 
and will decrease with increasing $x$ and $m_2$.  We also expect that our expression for $x$ will be an 
overestimation, since we have assumed that the most popular $m_1$ strategy is ranked as low as it possibly can 
be in the $m_2$ RSS.  We believe that these two factors explain why our theoretical predictions for the 
standard deviation in the excess demand slightly underestimate the numerical results for the multiple-memory 
populations.  
We can therefore conclude that multiple-memory populations gain their comparative advantage by behaving as 
mixed-ability populations with fewer strategies.  Additional strategies in the multiple-memory populations 
will cause the standard deviation in the excess demand to approach the random limit. However, the rate is far 
slower than in the case of  either pure populations or mixed-ability populations.

In game realizations where both the $m_1$ RSS and $m_2$ RSS are not crowded (e.g. as in the $s=3$ 
multiple-memory populations in Figs. 2 and 3) our simple theory for $x$ does not hold.  In cases where one of 
the RSS is not crowded, the highest ranked $m_1$ strategy can map to a higher-ranked $m_2$ strategy than we 
had assumed above.  This causes the value for $p$ to be reduced, and could in certain cases cause $x$ to fall 
below $s_2/s$ as in the $s=3$ case above.  We suspect that this effect is related to the information in the 
history string.  If the $m_1$ agents can fully access the information in the $m_1$ length bit-string, but 
there are insufficient $m_2$ strategies to access the additional information in the $m_2$ length bit-strings, 
it may be more advantageous to play an $m_1$ strategy.  This conjecture is reinforced by the importance of 
memory in the multiple-memory populations, which we believe is related to the different time scales being 
tracked by the system.  When neither of the RSS are crowded, we expect $x \sim s_2/s$. (This result has been 
confirmed in numerical simulations with $m_1=10$ and $m_2=13$).

In conclusion, we have studied the performance and dynamics of a population of multiple-memory agents 
competing in an artificial market.  We have shown that multiple-memory agents possess a comparative advantage 
over both pure populations of agents and mixed-ability populations.  We have presented a theory based on the 
Crowd-Anticrowd theory, which is in good agreement with these numerical results.

\acknowledgments

KEM is grateful to the Marshall Aid Commemoration Commission for support.

\end{document}